\documentclass[
preprint,
 amsmath,amssymb,
pra,
]{revtex4-1}
\usepackage{dcolumn}
\usepackage{bm}
\usepackage{amsfonts}
\usepackage{amsmath}
\usepackage{amssymb}
\usepackage{float}
\usepackage{graphicx}
\usepackage{subfigure}
\usepackage{epstopdf}
\usepackage{hyperref}

\begin{document}
\title{Dissipative generation of significant amount of photon-phonon asymmetric steering in magnomechanical interfaces}
\author{Tian-Ang Zheng$^{1}$}
\author{Ye Zheng$^{1}$}
\author{Lei Wang$^{1}$}
\author{Chang-Geng Liao$^{1}$}
\thanks{cgliao@zjgsu.edu.cn}
\affiliation{$^{1}$ School of Information and Electronic Engineering (Sussex Artificial Intelligence Institute), Zhejiang Gongshang University, Hangzhou, 310018, China}

\date{\today}
\begin{abstract}
  We propose an effective approach for generating significant amount of entanglement and asymmetric steering between photon and phonon in a cavity magnomechanical system which consists of a microwave cavity and a yttrium iron garnet sphere. By driving the magnon mode of the yttrium iron garnet sphere with blue-detuned microwave field, the magnon mode can be acted as an engineered resevoir cools the Bogoliubov modes of microwave cavity mode and mechanical mode via beam-splitter-like interaction. In this way, the microwave cavity mode and mechanical mode are driven to two-mode squeezed states in the stationary limit. In particular, strong two-way and one-way asymmetric quantum steering between the photon and phonon modes can be obtained with even equal dissipation. It is very different from the conventional proposal of asymmetric quantum steering, where additional unbalanced losses or noises on the two subsystems has been imposed. Our finding may be significant to expand our understanding of the essential physics of asymmetric steering and extend the potential application of the cavity spintronics to device-independent quantum key distribution.
\end{abstract}

\maketitle

\section{Introduction}
Hybridizing distinct physical systems possessing complementary charateristics is crucial for practical quantum information applications. In the last few years, hybrid quantum systems based on magnonics has attracted extensive attention and gradually developed as a promising platform for quantum information processing~\cite{RogersLoGulloDeChiaraPalmaPaternostro+2014}. The quanta of collective spin excitations in ferromagnetic crystals, i.e., magnon, can coherently coupled to microwave and optical photons, as well as phonons through magnetic dipole~\cite{PhysRevLett.111.127003,PhysRevLett.113.083603,PhysRevApplied.2.054002,PhysRevLett.113.156401,TABUCHI2016729}, magneto-optical~\cite{PhysRevB.93.174427,PhysRevLett.116.223601,PhysRevLett.117.123605,PhysRevLett.117.133602,PhysRevA.94.033821,PhysRevB.96.094412,Osada_2018,PhysRevLett.120.133602}, and magnetostrictive interactions~\cite{zhang2016cavity,PhysRevLett.121.203601} respectively. Besides, coherent effective
interaction between the magnon and superconducting qubit in the cavity can be realized via the coupling of them to the same cavity modes~\cite{lachance2017resolving} or the virtual photon excitation~\cite{TABUCHI2016729,tabuchi2015coherent,2019arXiv190312498W}. The successful development of the system has attracted considerable interest into this field. A variety of novel phenomena have been explored, such as magnon gradient memory~\cite{zhang2015magnon}, level attraction~\cite{PhysRevB.98.024406,PhysRevB.99.134426,PhysRevLett.121.137203,PhysRevLett.123.227201}, exceptional points~\cite{zhang2017observation,PhysRevLett.123.237202,PhysRevLett.125.147202}, manipulation of Distant Spin Currents~\cite{PhysRevLett.118.217201}, bistability~\cite{PhysRevLett.120.057202}, nonreciprocity~\cite{PhysRevLett.123.127202}, magnon laser~\cite{Liu:20}, etc.

Another attraction of cavity magnomechanical system is that it enables the exhibition of macroscopic quantum effects from the fundamental perspective. Therefore, a wide range of interests have been engaged to the generation of different types of macroscopic nonclassical states, including genuine tripartite~\cite{PhysRevLett.121.203601} or bipartite entanglement between subsystems~\cite{caiannalen,Li_2019,PhysRevLett.124.213604,PhysRevResearch.1.023021,PhysRevB.101.014419,PhysRevB.101.054402,nair2020deterministic,yu2020macroscopic,luo2021nonlocal,andp.202070031,wu2021remote}, squeezing of magnon and phonon~\cite{PhysRevResearch.3.023126,Zhang:21}, magnon Fock states~\cite{PhysRevA.100.013810}, among others. Recently, the proposal of optomagnonic Bell test is also presented~\cite{xie2021proposal}. Apart from to the studies mentioned above, considerable attention has been devoted to the generation of quantum steering~\cite{PhysRevApplied.15.024042,DING2021126903,zheng2021enhanced,PhysRevResearch.3.013192,PhysRevA.103.053712}, which is intrinsically distinct from quantum entanglement and Bell nonlocality for its asymmetric characteristics between the parties involved.

Here we propose an effective approach for generating significant amount of entanglement and asymmetric steering between photon and phonon in a cavity magnomechanical system which consists of a microwave cavity mode, macroscopic magnon and phonon modes of a yttrium iron garnet sphere. Light-mechanical quantum steering~\cite{PhysRevA.88.052121,PhysRevA.90.043805,PhysRevA.95.053842} or steering between two massive mechanical oscillators~\cite{PhysRevA.92.012126,PhysRevA.98.042115,PhysRevA.101.032120} have been widely studied in cavity optomechanical systems, suggesting that photon-phonon or phonon-phonon one-way quantum steering can be achieved in such systems~\cite{PhysRevA.95.053842,PhysRevA.98.042115,PhysRevA.101.032120}. Primary researches also indicate that asymmetric steering between two magnons~\cite{PhysRevApplied.15.024042,zheng2021enhanced} can be obtained in cavity magnonics and asymmetric steering between a macroscopic mechanical oscillator and a magnon mode~\cite{PhysRevResearch.3.013192} can be obtained in a microwave-mediated phonon-magnon
interface. Nevertheless, whether asymmetric steering can be directly obtained in cavity magnomechanical system is still almost in blank. Recently, by introducing a gain cavity mode, the authors~\cite{DING2021126903} show that magnon $\rightarrow$ phonon steering in $PT$-symmetry system can be significantly enhanced versus the case in the conventional cavity magnomechanical system, which opens up a route to explore the characteristics of steering in magnomechanical systems. Unlike the proposal in Ref.~\cite{DING2021126903}, which mainly focuses on the enhancement of magnon $\rightarrow$ phonon steering in $PT$-symmetry system, we show that significant amount of entanglement and asymmetric steering between photon and phonon can be directly obtained in a cavity magnomechanical system. By driving the magnon mode of the yttrium iron garnet sphere with blue-detuned microwave field, the magnon mode can be acted as an engineered resevoir cools the Bogoliubov modes of microwave cavity mode and mechanical mode via beam-splitter-like interaction. In this way, the microwave cavity mode and mechanical mode are driven to two-mode squeezed states in the stationary limit. In particular, strong two-way and one-way asymmetric quantum steering between the photon and phonon modes can be obtained with even equal dissipation. It is very different from the conventional proposal of asymmetric quantum steering, where additional unbalanced losses or noises on the two subsystems has been imposed. Our finding may be significant to expand our understanding of the essential physics of asymmetric steering as well as to extend the potential application of the cavity spintronics to device-independent quantum key distribution.
\section{the model and dynamics}\label{sec2}
The cavity magnomechanical system consists of a microwave cavity and a highly polished single-crystal yttrium iron garnet (YIG) sphere. The magnetic dipole interaction mediates the coupling between the microwave cavity mode $a$ and the collective motion of a large number of spins in YIG sphere, namely magnons $m$. The YIG sphere is also an excellent mechanical resonator $b$, which couple to magnons via magnetostrictive interaction. Therefore, the system is described by a radiation pressure-like Hamiltonian similar to cavity optomechanical system given by ($\hbar=1$)~\cite{PhysRevLett.121.203601}
\begin{eqnarray}
H& =&  \omega _a a^ \dag a+\omega_m m^ \dag m+\omega_b b^ \dag b+g(a+a^ \dag)(m+m^\dag)+\eta m^ \dag m(b+b^\dag)\nonumber\\
&&+ i E(t)(e^{-i\omega_dt}m^ \dag  -e^{i\omega_dt}m),
\end{eqnarray}
where $w_j$ represents the resonance frequency of the bosonnics mode $j$ ($j=a, b$, and $m$), $g$ and $\eta$ respectively denotes the magnetic dipole interaction and magnetostrictive interaction strength. The last term is the driving Hamiltonian of the magnon mode. The magnon frequency $w_m=\beta H_B$ is adjustable in a large range by altering the strength of the external bias magnetic field $H_B$ with respect to the gyromagnetic ratio $\beta$. The magnetic dipole interaction $g\propto \sqrt{N}$ with $N$ being the number of spins. Strong cavity-magnon coupling can be realized in recent experiments while the magnetostrictive interaction strength $\eta$ is typically weak. Nevertheless, the magnetostrictive interaction can be effectively enhanced by strong driving. The magnetic dipole interaction term $g(a+a^ \dag)(m+m^\dag)$ can be approximated into $g(am^\dag+a^ \dag m)$ by adopting the rotating-wave approximation.
Then, in a rotating frame with respect to $\omega_{d} (m^ \dag m+a^ \dag a)$, we obtain
\begin{eqnarray}\label{HR}
H_R& =&  \Delta_a a^ \dag a+\Delta_m m^ \dag m+\omega_b b^ \dag b+g(am^\dag+a^ \dag m)+\eta m^ \dag m(b+b^\dag)\nonumber\\
&&+ i E(t)(m^ \dag  -m),
\end{eqnarray}
with $\Delta_a=\omega _{a}-\omega_{d}$ and $\Delta_m=\omega _{m}-\omega_{d}$.

In noisy environments, the system dynamics is governed by the Heisenberg-Langevin equations (HLEs) \cite{gardiner2004quantum}:
\begin{subequations}\label{L}
\begin{align}
\dot{a}=&-(\kappa_a/2+i\Delta_a)a-ig m +\sqrt{\kappa_{a}}a ^{\rm{in}}(t),\\
\dot{b}=&-(\gamma/2+i\omega_b)b-i\eta m^ \dag m+\sqrt{\gamma}b^{\rm{in}}(t)\\
\dot{m}=&-(\kappa_m/2+i\Delta_m)m-iga-i\eta m (b+b^ \dag)+E(t)+\sqrt{\kappa_{m}}m ^{\rm{in}}(t).
\end{align}
\end{subequations}
Here, $\kappa_{a}$, $\gamma$, and $\kappa_{m}$ stand for the dissipation rates of the cavity mode, the phonon mode, and the magnon mode, respectively. $a^{\rm{in}}(t)$, $b^{\rm{in}}(t)$, and $m^{\rm{in}}(t)$ are input noise operators of photon, phonon, and magnon modes, and their auto-correlation functions satisfy the relation:
\begin{subequations}\label{auto}
\begin{align}
&\langle {{a}^{\rm{in}}(t)}{a^{\rm{in}\dag}(t')}\rangle  = \delta (t - t'),\\
&\langle {{a}^{\rm{in}\dag}(t)}{a^{\rm{in}}(t')} \rangle = 0,\\
&\langle {{b}^{\rm{in}}(t)}{b^{\rm{in}\dag}(t')}\rangle  = (\bar n_b+1)\delta (t - t'),\\
&\langle {{b}^{\rm{in}\dag}(t)}{b^{\rm{in}}(t')} \rangle = \bar n_b\delta (t - t'),\\
&\langle {{m}^{\rm{in}}(t)}{m^{\rm{in}\dag}(t')}\rangle  = \delta (t - t'),\\
&\langle {{m}^{\rm{in}\dag}(t)}{m^{\rm{in}}(t')} \rangle = 0,
\end{align}
\end{subequations}
with $\bar n_{b}=(\exp(\hbar \omega_b/(k_BT))-1)^{-1}$ being the mean thermal phonon number at temperature $T$.

In the presence of strong driving field, Eq. (\ref{HR}) can be linearized by substituting each system operators $o$ ($o=a, m$, and $b$) with the sum of steady-state mean values $\langle o(t)\rangle$ and quantum fluctuations $\delta o$, leading to the linearized Hamiltonian
\begin{eqnarray}\label{HL}
{H^{{\rm{lin}}}} & =& \Delta_a \delta a^\dag \delta a + \tilde{\Delta}_m \delta m^\dag \delta m +\omega_b \delta b^ \dag \delta b+g(\delta a\delta m^\dag+\delta a^\dag \delta m)\nonumber\\
&&+(G(t)^*\delta m+G(t) \delta m^\dag)(\delta b+\delta b^\dag),
\end{eqnarray}
where the effective detuning $\tilde{\Delta}_m=\Delta_m +\eta(\langle b(t)\rangle+\langle b(t)\rangle^*)\simeq \Delta_m$ and the effective coupling $G(t)=\eta\langle m(t)\rangle$.

After standard linearization techniques~\cite{PhysRevLett.98.030405} are applied to Eq. (\ref{L}), we obtain a set of differential equations for the steady-state mean values:
\begin{subequations}\label{mean}
\begin{align}
\dot{\langle a(t)\rangle}=&-(\kappa_{a}/2+i\Delta_a)\langle a(t)\rangle-ig\langle m(t)\rangle,\\
\dot{\langle b(t)\rangle}=&-(\gamma/2+i\omega_b)\langle b(t)\rangle-i\eta |\langle m(t)\rangle|^2,\\
\dot{\langle m(t)\rangle}=&-(\kappa_{m}/2+i\Delta_m)\langle m(t)\rangle-ig\langle a(t)\rangle-i\eta \langle m(t)\rangle(\langle b(t)\rangle+\langle b(t)\rangle^*)+E(t)
\end{align}
\end{subequations}
and the linearized QLEs for the quantum fluctuation operators:
\begin{subequations}\label{flu}
\begin{align}
\dot{\delta a}=&-(\kappa_{a}/2+i\Delta_a)\delta a-ig \delta m+ \sqrt{\kappa_{a}} a^{\rm{in}}(t),\\
\dot{\delta b}=&-(\gamma/2+i\omega_b)\delta b+i\eta (\langle m(t)\rangle^* \delta m+\langle m(t)\rangle \delta m^ \dag ) +\sqrt{\gamma}b^{\rm{in}}(t),\\
\dot{\delta m}=&-(\kappa_{m}/2+i\Delta_m)\delta m-ig \delta a-i\eta [(\langle m\rangle (\delta b+\delta b^{\dag})+\delta m(\langle b(t)\rangle+\langle b(t)\rangle^*) ]+ \sqrt{\kappa_{m}}m^{\rm{in}}(t).
\end{align}
\end{subequations}

By applying the blue-detuned ($\omega_1=\tilde{\Delta}_m+\omega_b$) and red-detuned ($\omega_2=\tilde{\Delta}_m-\omega_b$) two-tone driving lasers
\begin{eqnarray}\label{driving}
E(t)=\sum\limits_{j = 1,2}E_{j}e^{-i\omega_{j}t},
\end{eqnarray}
the asymptotic solution $\langle m(t)\rangle$ can be expressed as
\begin{eqnarray}
\langle m(t)\rangle\approx\sum\limits_{j = 1,2}\frac{E_{j}e^{-i\omega_{j}t}}{\kappa_{m}/2+i(\tilde{\Delta}_m\mp\omega_{j})+g^2/[\kappa_{a}/2+i(\Delta_a\mp\omega_{j})]}.
\end{eqnarray}
In the asymptotic regime, the Hamiltonian of Eq. (\ref{HL}) in the interaction picture through the unitary operator $U(t)=\exp[-it(\Delta_a \delta a^\dag \delta a + \tilde{\Delta}_m \delta m^\dag \delta m +\omega_b \delta b^ \dag \delta b)]$ becomes
\begin{eqnarray}\label{Hasy}
{H^{{\rm{asy}}}} & =& g\delta a\delta m^\dag e^{-i(\Delta_a-\tilde{\Delta}_m)t}+G_1\delta m^\dag\delta b^ \dag
+G_1\delta m^\dag\delta b e^{-2i\omega_bt}+G_2\delta m^\dag\delta b^ \dag e^{2i\omega_bt}\nonumber\\
&&+G_2\delta m^\dag\delta b +h.c.,
\end{eqnarray}
with
\begin{subequations}\label{Geff}
\begin{align}
G_1=&\frac{\eta E_1}{\kappa_{m}/2+i(\tilde{\Delta}_m-\omega_{1})+g^2/[\kappa_{a}/2+i(\Delta_a-\omega_{1})]}\\
G_2=&\frac{\eta E_2}{\kappa_{m}/2+i(\tilde{\Delta}_m+\omega_{2})+g^2/[\kappa_{a}/2+i(\Delta_a+\omega_{2})]}.
\end{align}
\end{subequations}
If we set $\Delta a =\tilde{\Delta}_m$, under the condition of weak coupling ( i.e., $G_1,G_2\ll \omega_b$), Eq. (\ref{Hasy}) becomes
\begin{eqnarray}\label{Heff}
{H^{{\rm{eff}}}} & =& g\delta a\delta m^\dag +G_1\delta m^\dag\delta b^ \dag+G_2\delta m^\dag\delta b +h.c..
\end{eqnarray}
In the following, we will respectively discuss the cases of two-tone driving and blue-detuned driving only.
Introducing three annihilation operators of Bogoliubove-mode
\begin{subequations}
\begin{align}
\beta_1=&S(r_1)\delta bS^\dag(r_1)=\delta b \cosh r_1+\delta b^\dag \sinh r_1,\\
\beta_2=&S(r_2)\delta aS^\dag(r_2)=\delta a \cosh r_2+\delta b^\dag \sinh r_2,\\
\beta_3=&S(r_2)\delta bS^\dag(r_2)=\delta b \cosh r_2+\delta a^\dag \sinh r_2,
\end{align}
\end{subequations}
where the squeezing operators $S(r_j)$ and squeezing parameters $r_j$ are defined as
\begin{subequations}
\begin{align}
S(r_1)=&\exp[r_1(\delta b\delta b-\delta b^\dag \delta b^\dag)],\\
S(r_2)=&\exp[r_2(\delta a\delta b-\delta a^\dag \delta b^\dag)],\\
r_1=&\tanh^{-1}(G_1/G_2),\\
r_2=&\tanh^{-1}(G_1/g).
\end{align}
\end{subequations}
The Hamiltonian of Eq.~(\ref{Heff}) is thus given by
\begin{eqnarray}\label{Heff1}
{H_1^{{\rm{eff}}}} & =& g\delta a\delta m^\dag +\tilde{G_1}\delta m^\dag\beta_1 +h.c.,
\end{eqnarray}
or
\begin{eqnarray}\label{Heff2}
{H_2^{{\rm{eff}}}} & =& G_2\delta b\delta m^\dag +\tilde{G_2}\delta m^\dag\beta_2 +h.c.,
\end{eqnarray}
with the coupling strengths
\begin{subequations}
\begin{align}
\tilde{G_1}=&\sqrt{G^2_2-G_1^2},\\
\tilde{G_2}=&\sqrt{g^2-G_1^2}.
\end{align}
\end{subequations}
Both Eqs.~(\ref{Heff1}) and~(\ref{Heff2}) are beam-splitter-like interaction Hamiltonian, which is well known
from optomechanical sideband cooling~\cite{PhysRevLett.99.093901,PhysRevLett.99.093902}. Under appropriate system parameters, the  Bogoliubove-mode $\beta_1$ or $\beta_2$ can be cooled to near ground state, generating single-mode squeezed state of
the mechanical mode $\delta b$ or two-mode squeezed state of the cavity mode $\delta a$ and mechanical mode $\delta b$. We note that the scheme to generate single-mode squeezed state of the mechanical mode $\delta b$ have been proposed in a cavity magnomehanical system by driving the magnon mode with red-detuned and blue-detuned microwave fields~\cite{Zhang:21}. The physical essence of the scheme is to adopt the cascaded dissipative
cooling process. Therefore, we will focus on the generation of two-mode squeezed state between the cavity mode $\delta a$ and mechanical mode $\delta b$ by driving the magnon mode with blue-detuned microwave field only.

\section{Evolution equation of the covariance matrix}\label{sec3}
Introducing column vectors of dimensionless quadrature operators and their input noises
\begin{subequations}
\begin{align}
R=&(q_{\delta a},p_{\delta a},q_{\delta b},p_{\delta b},q_{\delta m},p_{\delta m})^T,\\
N(t)=&(\sqrt{\kappa_{a}}q_{a^{in}},\sqrt{\kappa_{a}}p_{a^{in}},\sqrt{\gamma}q_{b^{in}},\sqrt{\gamma}p_{b^{in}},\sqrt{\kappa_{m}}q_{m^{in}},\sqrt{\kappa_{m}}p_{m^{in}})^T,
\end{align}
\end{subequations}
which are related to bosonic modes $o$ ($o\in \{\delta a,\delta b,\delta m,a^{\rm{in}}(t),b^{\rm{in}}(t),m^{\rm{in}}(t)\}$) via~$q_o=(o+o^\dag)/\sqrt{2}$ and $p_o=(o-o^\dag)/(i\sqrt{2})$, Eq.~(\ref{flu}) can be written in a compact matrix form$\colon$
\begin{equation}\label{dR}
\dot{R}=MR+N,
\end{equation}
with
\begin{eqnarray}\label{M}
{\footnotesize M = \left( {\begin{array}{*{20}{c}}
 -\kappa_{a}/2&0&0&0&0&g\\
0&-\kappa_{a}/2&0&0&-g&0\\
0&0&-\gamma/2&0&f_{1I}+f_{2I}&f_{1R}-f_{2R}\\
0&0&0&-\gamma/2&-f_{1R}-f_{2R}&f_{1I}-f_{2I}\\
0&g&f_{2I}-f_{1I}&f_{1R}-f_{2R}&-\kappa_{m}/2&0\\
-g&0&-f_{1R}-f_{2R}&-f_{1I}-f_{2I}&0&-\kappa_{m}/2\\
\end{array}} \right),}
\end{eqnarray}
where we have used the asymptotic approximation Hamiltonian of Eq. (\ref{Hasy}), and $f_{jR}$ and $f_{jR}$ are independent the real parts and imaginary parts of $f_{j}$ defined by
\begin{subequations}
\begin{align}
f_1(t)=&G_2+G_1e^{-2i\omega_bt},\\
f_2(t)=&G_1+G_2e^{2i\omega_bt}.
\end{align}
\end{subequations}
When focus only on blue detuning laser driving, $G_2=0$ in all the terms.

The linearized Hamiltonian of the system ensures that an initial Gaussian state will remains Gaussian~\cite{weedbrook2012gaussian} as the system is stable. To obtain the information-related properties of a Gaussian state, we introduce the $6\times6$ covariance matrix (CM) $\sigma$ with components defined by~\cite{adesso2007entanglement,olivares2012quantum,weedbrook2012gaussian}
\begin{eqnarray}\label{sigma}
\sigma_{j,k}= &\langle R_jR_k+R_kR_j\rangle /2.
\end{eqnarray}
Here, $R_j$ is the $j$th element of column vector $R$.
From Eqs. (\ref{auto}), (\ref{dR}), and (\ref{sigma}), it can be deduced \cite{Mari2009Gently}
\begin{equation}\label{dif}
\dot{\sigma}=M\sigma+\sigma M^T+D,
\end{equation}
where $D$ is a diffusion matrix with its components relevant to the noise correlation functions in Eq. (\ref{auto}) and defined as
\begin{equation}
D_{j,k}\delta(t-t')=\langle N_j(t)N_k(t')+N_k(t')N_j(t) \rangle/2.
\end{equation}
Actually, one can find that $D$ is diagonal
\begin{eqnarray}
D=\rm{diag}[\kappa_{a}/2,\kappa_{a}/2,\kappa_{b}(2\bar n_{b}+1)/2,\kappa_{b}(2\bar n_{b}+1)/2,\kappa_{m}/2,\kappa_{m}/2].
\end{eqnarray}
In the following, Eq.~(\ref{dif}) will be utilized to numerically simulate the time evolution of the entanglement and quantum steering between the photon and phonon modes.

For a Gaussian state of the two modes of interest, the logarithmic negativity $E_{\rm{N}}$ \cite{vidal2002computable,adesso2004extremal} is used to gauge its level of entanglement. As for quantum steering, a computable criterion based on the form of quantum coherent information have been introduced \cite{PhysRevLett.114.060403}. Both of the above measures can be computed from the reduced $4\times4$ CM ${\sigma_R}(t)$ for the photon and phonon modes
\begin{equation}
{\sigma_R}(t) = \left( {\begin{array}{*{20}{c}}
{{\sigma _1}}&{{\sigma _{\rm{3}}}}\\
{\sigma _{\rm{3}}^{\rm{T}}}&{{\sigma _2}}
\end{array}} \right),
\end{equation}
with each $\sigma_j$ being a $2\times2$ subblock matrices of ${\sigma_R}(t)$. The entanglement is then calculated by
\begin{equation}
E_{\rm{N}} = \max [0, - \ln (2\vartheta)]
\end{equation}
with
\begin{equation}
\vartheta \equiv 2^{-1/2}\{\Sigma_--[\Sigma_-^2-4I_4]^{1/2}\}^{1/2}
\end{equation}
and
\begin{equation}
\Sigma_- \equiv I_1 + I_2 - 2I_3,
\end{equation}
where $I_1=\det{\sigma _1}$, $I_2=\det{\sigma _2}$, $I_3=\det{\sigma _{\rm{3}}}$, and $I_4=\det{\sigma_R}$ are the symplectic invariants.
The level of steerability from photon to phonon is given by
\begin{equation}
G_A\equiv G^{a\rightarrow b}({\sigma _{\rm{R}}}) = \max[0, \frac{1}{2} \ln\frac{I_1}{4I_4}],
\end{equation}
and a corresponding measure of Gaussian $b\rightarrow a$ steerability can be obtained by swapping the corresponding item of photon and phonon, resulting in an expression
\begin{equation}
G_B\equiv G^{b\rightarrow a}({\sigma _{\rm{R}}}) = \max[0, \frac{1}{2} \ln\frac{I_2}{4I_4}].
\end{equation}

\section{The results and discussion}\label{sec5}

\begin{figure}[t]\label{fig:1}
  \centering
  {\includegraphics[width=0.3\textwidth]{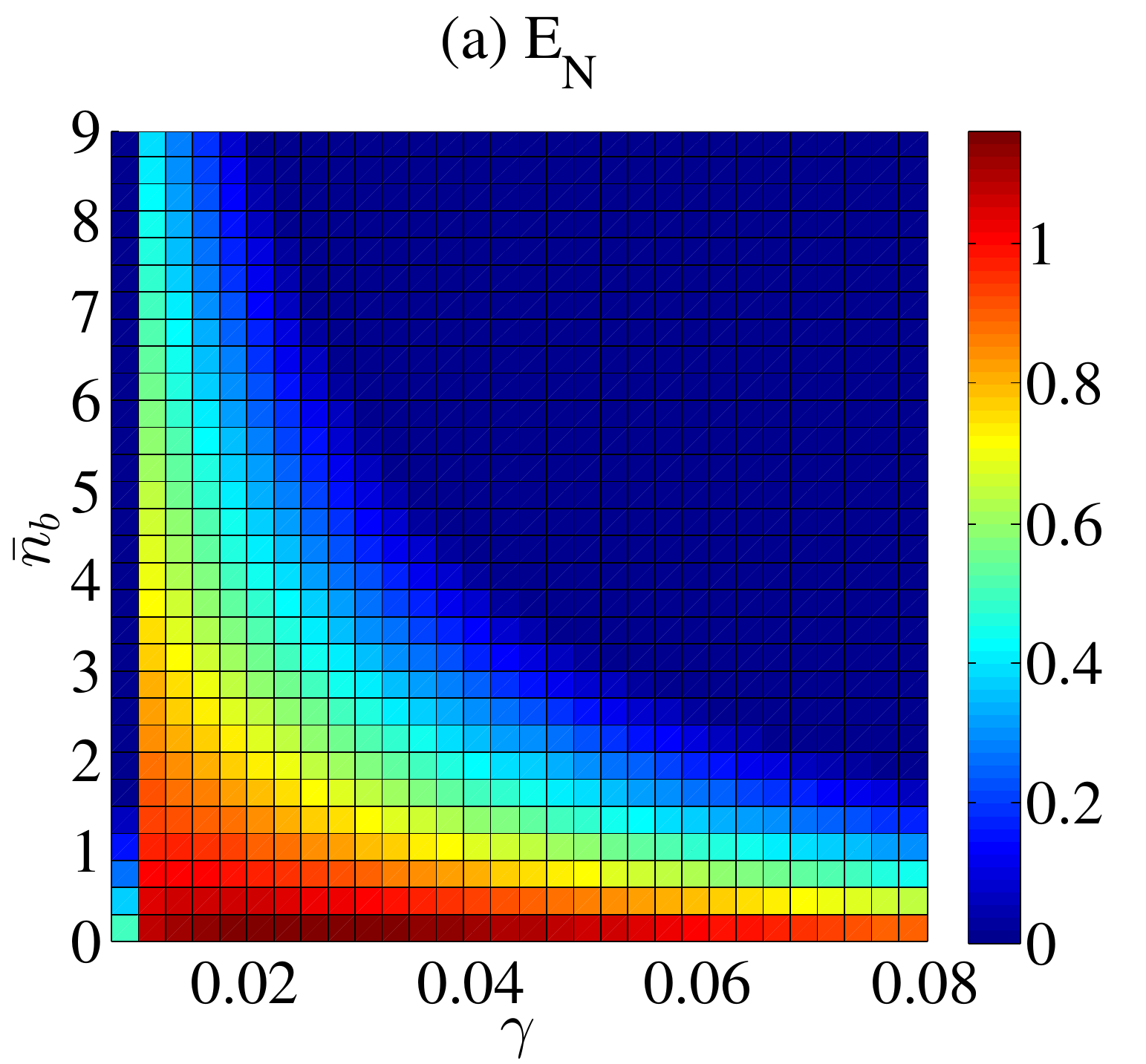}\label{fig:1(a)}
  \includegraphics[width=0.3\textwidth]{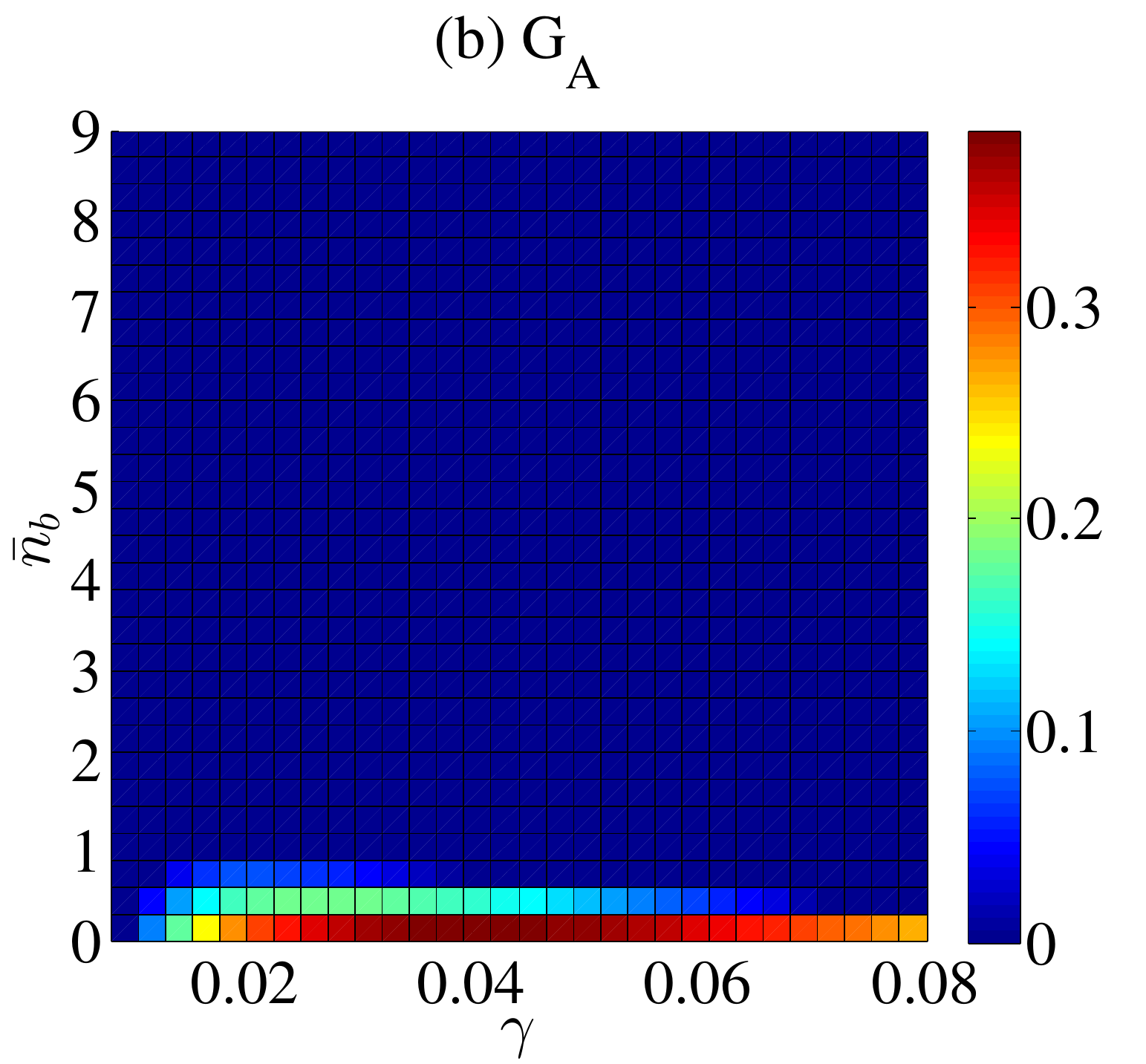}\label{fig:1(b)}
  \includegraphics[width=0.3\textwidth]{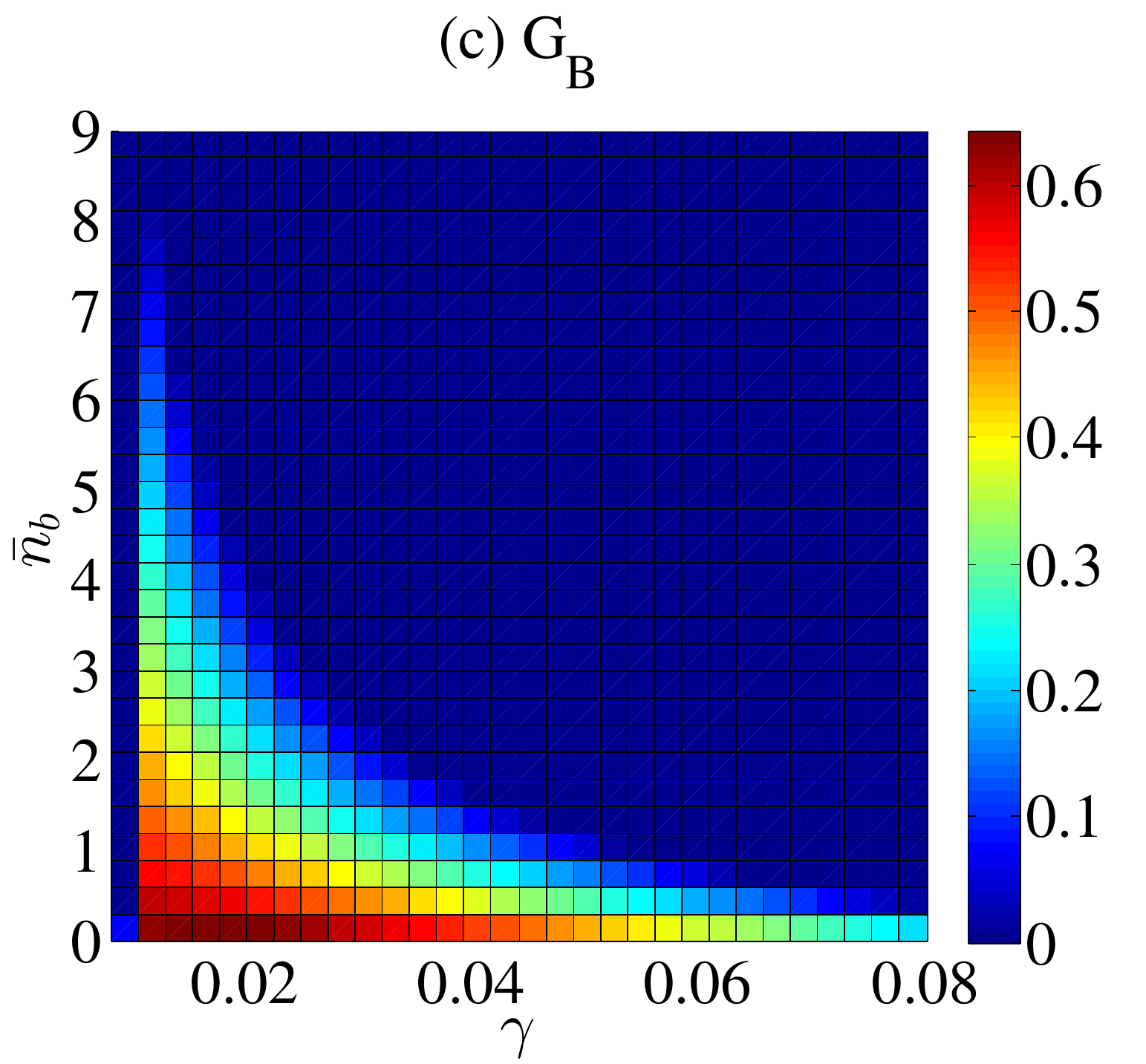}\label{fig:1(c)}}
  \caption{\label{fig:1}
  (Color online) Maximum entanglement and steering for each time period in the asymptotic regime as functions of $\gamma$ and $\bar{n}_b$. The chosen parameters in units of $\omega_b$ are: $\Delta_a=\Delta_m=1000$, $\kappa_{a}=0.02$, $\kappa_{m}=0.3$, $g=0.28$, $\eta=2\times10^{-8}$, $G_-=0.21$, and $G_+=0$.}
\end{figure}
In order to demonstrate the feasibility of our scheme, we plot in Figure~\ref{fig:1} the peak values of stationary entanglement and steering between the microwave cavity mode $\delta a$ and the mechanical mode $\delta b$ for each time period in the asymptotic regime as functions of mechanical dissipation rate $\gamma$ and mean thermal phonon numbe $\bar{n}_b$. On the one hand, it can be found from Fig.~\ref{fig:1} that the maximal values of entanglement and steering all decrease with the increase of $\bar{n}_b$. The phenomenon is obvious and easy to understand. When the mechanical dissipation rate $\gamma$ takes a fixed value, the interaction between the
mechanical mode $\delta b$ and its thermal bath accordingly maintains a constant intensity, the increase of $\bar{n}_b$ (i.e., raising the temperature $T$ of thermal bath) will raise the final effective temperature of Bogoliubov-mode $\beta_2$ (especially
the mechanical mode $\delta b$). Since both entanglement and quantum steering are sensitive to the environmental temperature, the optimal values of them gradually decrease with the increase of $\bar{n}_b$. However, due to the asymmetry of the system, the $a\rightarrow b$ steering $G_A$ drops faster than the entanglement $E_N$ and the $b\rightarrow a$ steering $G_B$. Therefore, there are states that only have
one-way quantum steering under properly parameters. Fig.~\ref{fig:1}, on the other hand, shows that the maximal values of entanglement and steering are nonmonotonic functions of $\gamma$ and take a maximum for a specific $\gamma$. Peculiarly, entanglement and steering are more sensitive to $\gamma$ with the increase of $\bar{n}_b$. The larger of $\gamma$ means stronger interaction between the
mechanical mode $\delta b$ and its thermal bath. When the mean thermal phonon number $\bar{n}_b$ is not negligible, the enlarged $\gamma$ will quickly raise the final effective temperature of Bogoliubov-mode $\beta_2$. Therefore, entanglement and steering drop quickly. When $\bar{n}_b$ is small, although increasing $\gamma$ will increase the interaction between the mechanical mode $\delta b$ and its thermal bath, the increase in effective temperature of $\beta_2$ is not significant. Accordingly, entanglement and steering are less affected by the variation of $\gamma$. From the above analyses, it seems that the smaller of the dissipation rates $\gamma$ of the phonon mode, the larger amount of entanglement and steering can be obtained. However, the mechanical resonator under larger dissipation has a smaller steady-state mean phonon number, which is one of the conditions for acquisition of the Hamiltonian of Eq.~(\ref{Heff}) (i.e., $\tilde{\Delta}_m=\Delta_m +\eta(\langle b(t)\rangle+\langle b(t)\rangle^*)\simeq \Delta_m$). Taking all of the above into account, the maximal values of entanglement and steering are nonmonotonic functions of $\gamma$ and take a maximum for a specific $\gamma$. The physical mechanism is somewhat similar to that of earlier study~\cite{PhysRevA.101.032120}.  As have been previously studied in Ref.~\cite{PhysRevA.101.032120}, the steerability between two mechanical modes is simultaneously affected both by the modes damping (loss) and thermal bath (noises), and the mechanical mode with larger damping rate is more difficult to be steered by the other one when the mean thermal occupancy of the mechanical baths are not negligible. Within the parameters that we are discussing about, $\kappa_a$ is set to be $0.02$. The $a\rightarrow b$ steering $G_A$ will be greater than the $b\rightarrow a$ steering $G_B$ only if both the conditions of larger $\gamma$ and smaller $\bar{n}_b$ are satisfied (e.g., $\gamma=0.07$ and $\bar{n}_b=0$), otherwise, $G_B\geq G_A$. The results enriches the earlier studies~\cite{PhysRevA.91.032121,PhysRevA.99.022335,Rosales-Zarate:15,obser2012}, where only the mode with larger dissipation rate can be steered by the other one. Besides, significant amount of two-way and one-way asymmetric quantum steering between the photon and phonon modes can be obtained with even equal dissipation (e.g., $\kappa_a=\gamma=0.02$). It is very different from the conventional proposal of asymmetric quantum
steering, where additional unbalanced losses or noises on the two subsystems has been imposed.

\section{Conclusions}\label{sec4}
In summary, we have proposed a theoretical scheme to achieve significant amount of steady-state entanglement and asymmetric steering between photon and phonon in a cavity magnomechanical system. By driving the magnon mode with blue-detuned microwave field, the magnon mode can be acted as an engineered resevoir cools the Bogoliubov modes of microwave cavity mode and mechanical mode via beam-splitter-like interaction. In this way, the microwave cavity mode and mechanical mode are driven to two-mode squeezed states with asymmetric steering in the stationary limit. The numerical simulation results reveal that the maximal values of entanglement and steering all decrease with the increase of $\bar{n}_b$, but are nonmonotonic functions of $\gamma$ and take a maximum for a specific $\gamma$. Peculiarly, entanglement and steering are more sensitive to $\gamma$ with the increase of $\bar{n}_b$. The underlying physical mechanism is analyzed in detail. Moreover, strong two-way and one-way asymmetric quantum steering between the photon and phonon modes can be obtained with
even equal dissipation. It is very different from the conventional proposal of asymmetric quantum steering, where additional unbalanced losses or noises on the two subsystems has been imposed. Our results may be significant to expand the understanding of the essential physics of asymmetric steering and extend the potential application of the cavity spintronics to device-independent quantum key distribution.

\section*{Acknowledgments}
 We acknowledge supports from the National Natural Science Foundation of China (Grants No.~12004336, No.~12075205, and No.~62071430) and funds from Zhejiang Gongshang University.

\bibliography{asymmetricsteering}
\bibliographystyle{apsrev4-1}

\end{document}